\documentclass[12pt]{article}
\pdfoutput=1
\usepackage{graphics, color}
\usepackage{graphicx}
\usepackage{amsmath}
\usepackage{amssymb}
\usepackage{amsfonts}
\usepackage{tensor}
\usepackage[usenames,dvipsnames]{xcolor}
\usepackage[normalem]{ulem}
\usepackage[bottom]{footmisc}

\definecolor{orange}{rgb}{1,0.5,0}
\definecolor{grey}{rgb}{.5,.5,.5}
\definecolor{bluegreen}{rgb}{0,.5,.5}
\definecolor{darkgreen}{rgb}{0,.5,0}

\def\gsim{\, \rlap{$>$}{\lower 1.1ex\hbox{$\sim$}}\,}
\def\lsim{\, \rlap{$<$}{\lower 1.1ex\hbox{$\sim$}}\,}
\newcommand{\be}{\begin{equation}}
\newcommand{\ee}{\end{equation}}
\newcommand{\bea}{\begin{eqnarray}}
\newcommand{\eea}{\end{eqnarray}}

\let\oldbibliography\thebibliography
\renewcommand{\thebibliography}[1]{%
  \oldbibliography{#1}%
  \setlength{\itemsep}{0pt}%
}
\begin{document}


\begin{titlepage}
\bigskip
\bigskip\bigskip\bigskip
\centerline{\Large \bf String theory to the rescue}
\bigskip


\bigskip\bigskip\bigskip
\bigskip\bigskip\bigskip

 \centerline
 {\bf Joseph Polchinski\footnote{\tt joep@kitp.ucsb.edu }}
 \medskip
 
 \centerline{\em Kavli Institute for Theoretical Physics}
\centerline{\em University of California}
\centerline{\em Santa Barbara, CA 93106-4030 USA}

\bigskip\bigskip\bigskip

\begin{abstract}
The search for a theory of quantum gravity faces two great challenges: the
incredibly small scales of the Planck length and time,
and the possibility that the observed constants of nature are in part the
result of random processes.  A priori, one
might have expected these to be insuperable obstacles.  However, clues
from observed physics, and the discovery of string theory,
raise the hope that the unification of quantum mechanics and general
relativity is within reach.

Prepared for the meeting ``Why Trust a Theory? Reconsidering Scientific Methodology in Light of Modern Physics,''  Munich, Dec. 7-9, 2015.
\end{abstract}
\end{titlepage}

\baselineskip = 16pt
\tableofcontents

\baselineskip = 18pt

\setcounter{footnote}{0}


\section{The Planck scale}

I am sorry to have to miss this meeting.  The question of how we are to understand quantum gravity is vitally important, and I would like to look at it in a broad way.

I will begin with 
the first, and arguably the most important, calculation in this subject.  This is Planck's 1899 use of dimensional analysis, combining the speed of light $c$, the gravitational constant $G$, and his own constant $\hbar$, to identify a fundamental length scale in nature~\cite{planck},
\be
l_{\rm P} = \sqrt{\hbar G / c^3} = 1.6 \times 10^{-33}\ {\rm cm} \,.  \label{planck}
\ee
Planck had a strong sense of the importance of what he had done:
\begin{quote}
These necessarily retain their meaning for all times and 
for all civilizations, even extraterrestrial and non-human ones, and can therefore be designated as {\it natural units}.
\end{quote}
I want to ask the question,
\begin{quote}
In light of this calculation, what would Planck have predicted for the state of fundamental physics today, 116 years later?
\end{quote}
Of course, prediction in science is hard, but the incredibly small scale of the Planck length is a powerful fact.
In Planck's time it had taken approximately 300 years to move {\it four} orders of magnitude, from the $10^{-4}$ cm achievable by optical microscopes to the $10^{-8}$ cm of the atomic scale.  Now he was arguing that to reach the natural length scale of nature one would have to improve by a further {\it twenty-five} orders of magnitude.
\begin{figure}[!h]
\begin{center}
\vspace {-5pt}
\includegraphics[scale=.6]{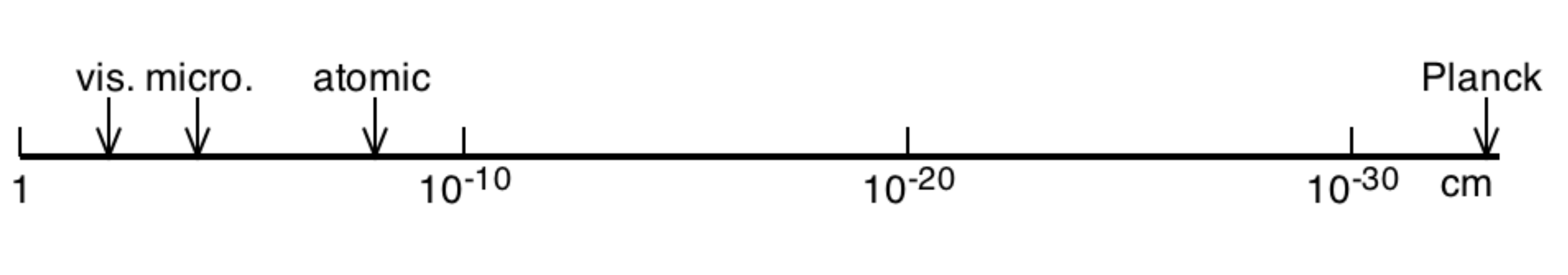}
\end{center}
\vspace {-10pt}
\caption{The gulf to be crossed.
}
\label{fig:radii}
\end{figure}

The first thing that Planck might have predicted is the existence of this meeting.  He could anticipate that this natural scale would remain inaccessible to direct observation for a very long time, perhaps forever.  So we would be confronted with the problem of how we are to proceed, and that is what this meeting is about.  

It is interesting to consider the situation if Planck's calculation had worked out to $10^{-17}$ cm.  This would still have seemed remote to him, nine orders of magnitude past the atomic scale.  But this is the length scale that the LHC is reaching today.  If this had been the Planck length, we would not be sitting around here whining about falsifiability, but instead anticipating a wonderland of experimental quantum gravity.  
\begin{figure}[!h]
\begin{center}
\vspace {-5pt}
\includegraphics[scale=.35]{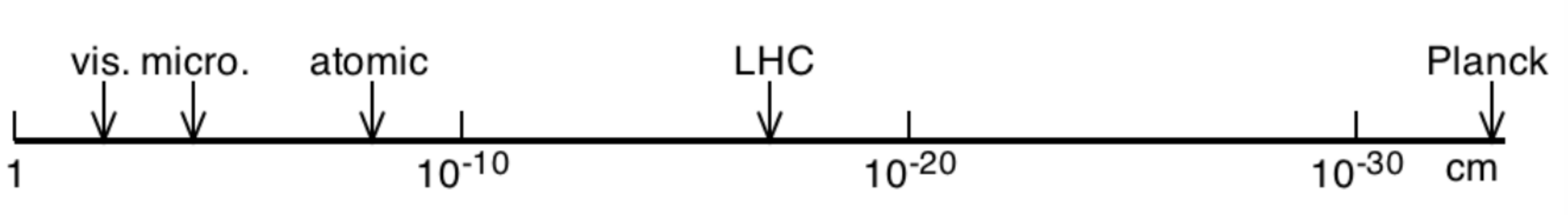}
\end{center}
\vspace {-10pt}
\caption{Progress to date.
}
\label{fig:radii2}
\end{figure}

I am a practical physicist.  I don't use expressions like `post-empirical.'  But physics is driven by numbers, and scales, and the fact that the Planck length is $10^{-33}$ cm and not $10^{-17}$ cm very much affects the tools that I can use, and the rate at which I can expect to make progress.  Theories in the making are judged in a different way than finished theories.  And because the Planck scale is so remote, this first period is likely to be much longer than we are used to.
For those who say that science has changed, what has changed is the magnitude of what we are trying to do, of what we need to do given what Nature has presented us with.

So how are we to proceed?  After moving four orders of magnitude in 300 years, and then nine orders in a little more than 100 years, we might continue to progress incrementally, and hope that some sort of scientific Moore's law will carry us the remaining sixteen orders of magnitude in time.  Certainly, there is no shortage of effort going into the next power of ten, and there are many indirect probes of higher energy physics, though almost all of these have been negative.  But there is reason to hope that we are in a position to leap to the answer. 

There have been times when theory has been able to leap a smaller gap: from Maxwell to light, from Dirac to antiparticles, from the Standard Model to the gluon, $W$, $Z$, top, and Higgs.  The gap here is vastly larger than any we have met before, but at the same time our theoretical sophistication has grown, and we have great theories, quantum mechanics and relativity, to build on. 

There is a danger of defining science too rigidly, so that one might decree that any discussion of the physics at $10^{-33}$ cm is unscientific because it is beyond reach of direct observation.  This was the attitude for a long period after Planck, lingering even when I was a graduate student.  But this makes science too weak, to decree that there are aspects of the natural world that are outside of its domain.  We have to make the effort; it may  prove to be fruitless, or premature, but I will argue that the situation is not so bad.

Let me continue my Planckian prediction game.  One scenario Planck might have imagined is that his calculation was overly pessimistic.  Dimensional analysis only works if one has identified all the relevant constants, and perhaps he had the wrong ones.  But the subsequent history seems to have borne out his choice.  Each of his constants, $c$, $G$, and $\hbar$, was about to launch a scientific revolution --- special relativity, general relativity, and quantum mechanics --- and these revolutions are still the center of our understanding of space, time, matter, and even reality.  Further, various ways in which quantum gravity might have manifested at longer distances --- large extra dimensions, low-scale strings, violations of Lorentz invariance (though this is problematic, as I will discuss later) --- have thus far failed to appear.  So we don't seem to be in the optimistic scenario.

At the opposite extreme, Planck  might have imagined that quantum gravity would largely be ignored, for how can science progress without observation?  Of course, given the importance of the problem, physicists would continue to dabble in it from time to time, and a few would devote themselves to it.   But this is not the present situation either.  

What has actually happened is something that would have been difficult to predict.  A large fraction of the theoretical physics community is working on quantum gravity, not with the enthusiasm that they would have had in the optimistic scenario, but certainly with far more enthusiasm and sense of progress than in the pessimistic scenario.  Something has happened, and that something is the discovery of string theory.

Of course, it is a sociological phenomenon that I have described so far, but it is the reflection of a scientific one.  But let me first analyze  the sociology a bit further.  One might think that  my claim is  circular: ``Of course, string theorists work on string theory.''  But science does not fall into such narrow silos.  Indeed, it is striking that, with few exceptions, those who have contributed major ideas to string theory have also made major contributions to other areas of science, including particle physics~\cite{Migdal:1966tq,Gross:1973id,Susskind:1978ms,GellMann:1980vs,Crewther:1979pi,Dine:1981rt,Goodman:1984dc,ArkaniHamed:1998rs,Randall:1999ee}, quantum field theory~\cite{Polyakov:1974ek,Banks:1981nn,Polchinski:1983gv,Witten:1988hf,Seiberg:1994pq,Maldacena:1997re}, mathematics~\cite{Candelas:1990rm,Witten:1994cg,Strominger:1996it,Connes:1997cr,Ooguri:2004zv}, condensed matter physics~\cite{Friedan:1983xq,Moore:1991ks,Polchinski:1992ed,Ryu:2006bv,Hartnoll:2007ih}, general relativity~\cite{Witten:1981mf,Garfinkle:1990qj,Damour:1994ya,
Gibbons:1997xz}, cosmology~\cite{Bousso:2000xa,Maldacena:2002vr,Kachru:2003aw,Alishahiha:2004eh}, nuclear physics~\cite{Kovtun:2004de,Liu:2006ug,Bhattacharyya:2008jc,Gubser:2010ze}, and most recently quantum information theory~\cite{Pastawski:2015qua,kitaev}.\footnote{I felt it important to illustrate this point with examples, but I have certainly omitted other scientists and papers of equal importance.  Most of these examples are physicists who are generally identified as string theorists contributing in other areas, but some (e.g.~\cite{ArkaniHamed:1998rs,Randall:1999ee,Kovtun:2004de,kitaev})
are physicists from other areas employing ideas from string theory, usually to the enrichment of both fields, while others are a combination of the two (e.g.\ \cite{Garfinkle:1990qj,Strominger:1996it,Connes:1997cr,Moore:1991ks,Damour:1994ya,Ryu:2006bv,Hartnoll:2007ih,Kachru:2003aw,Liu:2006ug}).}
 
There is a rather silly criticism of string theory~\cite{smolin}, that string theorists are not enough like Einstein.  This focuses on Einstein's philosophical bent.  But there was more to Einstein.  In addition to his work on relativity, Einstein explained the photoelectric effect in terms of the quantization of light, and Brownian motion in terms of atoms, predicted stimulated emission, and understood the heat capacity of matter, among many other insights.  It seems that there is a unity to physics, so that the ability to perceive new principles is not limited to narrow areas.  

I included this list to show that the scientists working on string theory are theoretical physicists in a broad sense (and many of those working on other approaches to quantum gravity suffer for a lack of this).  But this also helps to explain how string theory is able to remain vital even with limited data.  And, I am happy to see that many young people appear on the list.

\section{String theory}

I am arguing that the discovery of a theory like strings is a surprise, something that Planck could not  have anticipated.  To see why, I will enumerate five features: the solution to the short distance problem, the uniqueness of the dynamics, the unification of physics and geometry,  the duality between gauge fields and strings, and new insights into the quantum mechanics of black holes.

\subsection{The short distance problem} 

The first question to ask is, what do our existing theories predict for physics at shorter distances.

If we simply take Einstein's theory and feed it into the path integral, applying the standard recipe for quantizing a classical theory, we get infinities.  Again this is implicit in Planck's calculation~(\ref{planck}): setting $\hbar = c = 1$, the gravitational coupling $G$ has units of length (cm) squared.  If we probe the physics at some length scale $d$, the dimensionless coupling is $G/d^2$, growing stronger as $d \to 0$.  In the quantum theory there will be virtual effects running over all values of $d$, giving a divergent result even for observations on longer distance scales.  The typical observation would be the scattering of gravitons, or more geometrically the measurement of quantum corrections to the solutions to Einstein's equations.  In the language of particle physics this is the problem of nonrenormalizability; more geometrically it is the problem of spacetime foam, quantum fluctuations tearing spacetime apart at short distances.

This problem is not unique to gravity.  The Fermi theory of the weak interaction also has a coupling constant ($G_F$) with units of cm$^2$.  This was a key clue that allowed theorists to predict the $W$, $Z$, and higgs, and most of their detailed properties, before experiment gave any direct sign that they even existed --- one of the gaps that has successfully been leapt.

To see why this clue is so powerful, imagine that we try to make a new theory that solves this problem, by smearing out the interaction so that it is not so strong at short distances.  Because of special relativity, smearing in space implies smearing in time, so we lose causality or unitarity and our theory does not make sense.  It would be easy to cure the problem, and find many candidate theories of the weak interaction or of gravity, if we were willing to give up Lorentz invariance.  But Lorentz invariance has been confirmed to very high accuracy, and even it were broken only at very short distances, virtual effects would transmit large breaking to observed quantities~\cite{Collins:2004bp}.  So this is the first thing that Planck might not have anticipated, that our theories of physics would become so tightly intertwined as to allow few ways forward.

In the case of the weak interaction, introducing the $W$, $Z$, and higgs produced the necessary smearing in a physically sensible way.  In the case of gravity the problem is harder.  Rather than just adding a few particles, it seems necessary to change the nature of all particles, from points to strings.  At least, that was the first thing that worked.  

There are other ideas out there.  I have reservations about most or all of them (for example, many violate Lorentz invariance).  But this is not so relevant to my talk, because the thing I am trying to explain is not why theorists work on one theory of quantum gravity rather than another, but why they work on quantum gravity at all.  I am arguing that string theory is successful on an absolute scale, not a relative one.  This requires positive arguments, not negative ones, and so we move on to the next positive.

\subsection{Uniqueness of dynamics}

A remarkable feature of string theory is that the dynamics, the equation of motion, is completely fixed by general principle.  This is consistent with the overall direction of fundamental theory,  describing the vast range of phenomena that we see in terms of fewer and fewer underlying principles.  Uniqueness would seem to be the natural endpoint to this process, but such theories are truly rare.

General relativity for a time seemed to have this property, with the equivalence principle determining the form of the Einstein equation.  Of course, the cosmological constant was a confusion, and the equation does not describe matter, but more importantly the restriction to terms linear in the curvature was artificial.  The equivalence principle allows terms quadratic in the curvature, cubic, and so on.  Indeed, even if one tries to omit them, they will be generated by quantum effects~\cite{Donoghue:1994dn}.  The reason that we do not usually discuss them is that their effects are minuscule for dimensional reasons --- Planck's calculation again --- but as a matter of principle they will be there.

In quantum field theory we also have many choices: gauge symmetries, field content, and couplings, including higher derivative terms.  There seems to be nothing else like string theory.\footnote{Supergravity comes closest, but its symmetries still allow an infinite number of higher derivative terms, and it does not include matter.}  Its existence is a surprising discovery, and I emphasize the word discovery.  It is a mathematical-physical structure that exists, and we are discovering it .

Indeed, when I assert that the equations of string theory are fully determined by general principle, I must admit that we do not yet know the full form of the equations, or the ultimate principle.  String theory was discovered in an incomplete and approximate form, and our understanding has gradually deepened.  In fact, string theory is no longer about the strings; our best understanding would put the holographic principle closer to the center.  But the uniqueness was visible even in the first form: the condition of world-sheet conformal invariance $T^a\!_a =0$ implies Einstein's equation plus matter, and the coupling constant becomes a field.  This property continues to hold in AdS/CFT.

\subsection{Physics from geometry}

In general relativity, gravity comes from the curvature of spacetime.   Unity of physics would suggest that the other interactions arise in this way as well.  But we have used up the evident part of spacetime in accounting for gravity, so there must be more to it.  The uniqueness of string theory actually forces more on us, extra dimensions plus dual structures such as branes.  It seems that there is about the right amount of structure to describe the physics that we see.  In particular, our laws of physics arise from the geometry of the extra dimensions.  Understanding this geometry ties string theory to some of the most interesting questions in modern mathematics, and has shed new light on them, such as mirror symmetry.

Coming from a background in quantum field theory and particle physics, I find this is less remarkable than most of the other features on my list, but it accounts for some of the great interest in string theory.  The physicists who work on string theory come from a great variety of scientific backgrounds (consider the list [2-40]) and would be expected to have very different intuitions as to what a fundamental theory should look like, yet string theory appeals to all of them.

\subsection{Duality between gauge fields and strings}

A  surprising discovery in recent decades about the structure of quantum physics is that duality is a common property.  
The word `quantize' suggests a one-to-one correspondence: go from the classical theory to the quantum theory by quantizing (for example with the path integral), and go back again by taking the classical limit.  For a long time, quantum field theory in particular was thought about in this way.  But now we know that many rich quantum theories have multiple classical limits, and that these can look quite different from one another.

Most remarkable is the duality between gauge theory and gravity~\cite{Maldacena:1997re} (presaged in~\cite{'tHooft:1973jz,Polyakov:1987ez,Banks:1996vh}).   A theory of quantum gravity, with all its puzzles, can be obtained by quantizing a gauge theory, something that we are very familiar with.  Even QCD, the theory of the strong interaction, contains quantum gravity, albeit in an unfamiliar regime of high curvature.  The gravitational theory obtained by this duality is restricted to spacetimes with special boundary conditions, but within this boundary one can have rich dynamics, including the graviton scattering discussed above, change of the topology of spacetime, or the black hole thought experiments to be discussed below.  Of course, extending this to more interesting spacetimes is a key research direction.

Moreover, when quantum gravity emerges in this way, so does the rest of string theory.\footnote{To be precise, if we have a theory of quantum gravity in AdS, it defines formally a field theory (CFT) on its boundary.  But for all CFT's for which we have an independent construction (e.g. via a Lagrangian), the emergent theory is string theory.}  Indeed, this is our most complete description of string theory.  It reproduces the approximate forms that we know in various limits, and fills in the parameter space between.  Moreover, the strings themselves are emergent as noted above: the starting point is no longer a theory of strings, and the principle by which the whole theory emerges is now best understood as  holography.

Recognizing a common origin for diverse phenomena has been one of the ongoing successes of science.
Maxwell's understanding of light in terms of electricity and magnetism was one of the great steps forward.  The duality between gravity and gauge theory is equal to this in intellectual magnitude and surprise.

\subsection{Quantum mechanics of black holes}

Nonrenormalizabilty is the immediate problem that one encounters when one tries to quantize gravity, but there are others.  Quantum mechanics and general relativity seem to give different pictures of the black hole.  In general relativity the black hole is hairless.  In quantum theory, it has a temperature and entropy, and so should have a statistical mechanical description in terms of a definite number (the exponential of the Bekenstein-Hawking entropy) of microscopic states.  This is the entropy puzzle: what are these states?  Beyond this is the information paradox.  Black hole evaporation seems to destroy information.  To be precise, this means that pure states evolve into mixed states~\cite{Hawking:1976ra}, in contradiction to the ordinary laws of quantum mechanics.  This is problematic, but so are the alternatives. 

Strominger and Vafa showed that the microstates could be understood in string theory, in agreement with the Bekenstein-Hawking count~\cite{Strominger:1996sh}.  Moreover, gauge/gravity duality shows that black hole evaporation must take place within the ordinary framework of quantum mechanics, contradicting Hawking.  This ability to shed light on difficult decades-old puzzles that are far from its original motivation is another success of the theory.

It should be noted that the black hole puzzles largely drove the discovery of gauge/gravity duality, forcing theorists to understand better the relation between the dynamics of black branes and D-branes.  It seems that these puzzles continue will to be a fruitful source of insight and new ideas~\cite{Almheiri:2012rt}.

\subsection{Summary}

Of the five properties I have discussed, I regard three (2.1, 2.3, and 2.5) as very important, and two (2.2 and 2.4) as remarkable.  Finding all of them in a single theory tells me that we have been fortunate.  In spite of the great gulf that faces us, we have learned enough from what we can see to figure out what lies on the other side.

\section{The multiverse}

Two of the positive features on my list --- physics from geometry, and uniqueness of dynamics --- come with a dark side.  If the physics that we see is determined by the geometry of spacetime, what determines that geometry?  It should solve Einstein's equation, or something like it.  But how many solutions does Einstein's equation have that look something like our universe, with four large dimensions and the rest small and compact?  This question was first addressed by Calabi and Yau, who provided the framework to show that the number is large, perhaps billions, even if we restrict to spaces with only the metric and no other fields.\footnote{To be precise, each of these has moduli, and so is actually a continuous family of solutions (formally the number of solutions is billions times infinity).  Realistic solutions~\cite{Silverstein:2001xn,Kachru:2003aw} do not have moduli.}  A more complete count is very indefinite, with the number $10^{500}$ often used as a stand-in (this arises as a small integer raised to a large power coming from topology~\cite{Bousso:2000xa,Susskind:2003kw,Douglas:2006es}).\footnote{A recent paper~\cite{Taylor:2015xtz} suggests a much larger number.  The early paper~\cite{Lerche:1986cx} gave the number $10^{1500}$, but it is not clear to what extent these represent discrete points in a much smaller set of moduli spaces.}  Even Einstein encountered a small landscape, the radius of the Kaluza-Klein theory.  If he had known about the nuclear forces, he would been led to the billion or so solutions of Calabi and Yau. 
 
Thus, the laws that we see do not follow uniquely from the dynamics, but depend also on the specific solution.  This is the price we pay for getting physics from geometry, and for having unique dynamical laws.  Of course, one of the central features of physics is that simple equations can have many and complicated solutions.  But we did not expect, or want, our observed physical laws to depend on the solution.

So what determines what solution, and laws, we find around us?  One might think that this depends on the initial conditions, and so we need to develop a theory of these.  Indeed we do need a theory of the initial conditions, but it will probably do little to solve our problem, because quantum mechanics and relativity conspire to hide the past.  If we start in any solution with a positive vacuum energy it will inflate.  Quantum mechanics will cause small regions to tunnel into any of the other vacua, and the process can repeat indefinitely.  We end up with a vast universe, with all possible solutions realized in one place or another.  This has come to be called the multiverse.    The things that we are trying to explain, the observed laws of nature,  then do not  follow from the fundamental theory but vary from place to place; that is, they are environmental.

In fact, there is reason to believe that we live in just such a universe~\cite{Linde:2015edk}.  The Standard Model vacuum is a rich place, with zero point energies, Higgs fields, quark condensates, color fluctuations, and so on. Why then is there not an enormous vacuum energy?
Virtually everyone interested in fundamental theory has wracked their brain over this.  For very general reasons, all known theories that predict a definite value of the cosmological constant give either an enormous value, or exactly zero.  In the latter case this is accompanied by unbroken supersymmetry, which we do not have.  Theories that do not predict a definite value are of  two types.  Either they have free parameters so the cosmological constant can be set to any value at all (fine tuning), or they have a multiverse.  The former case seems unsatisfactory: ultimately we expect that there will be a theory with definite dynamics, and we will be in one of the other cases.  In the latter case, different values of the cosmological constant will be realized in different regions.  However, most of these regions will be boring, without interesting structure.  For structure to develop requires many degrees of freedom, large volume, and long times; these will be realized only in the rare regions where the cosmological constant is small.  This was analyzed by Weinberg~\cite{Weinberg:1988cp}, sharpening arguments by others~\cite{DavUnr.1981,Linde:1984ir,Sakharov:1984ir,Banks:1984cw}.  So the only known class of theories that have definite dynamics and are consistent with our observations of the universe are those that give a multiverse.

Before 1997, it was widely assumed that the cosmological constant was exactly zero for some reason, perhaps a symmetry, that had not yet been discovered.  The only framework that  predicted otherwise was the multiverse~\cite{Weinberg:1988cp}, since the cosmological constant need only be small for structure to develop, and an exact zero value would be a set of measure zero.  Thus the discovery of the vacuum energy in 1997 came as a great surprise to all but two classes of people: those who had paid attention to the data without theoretical prejudice (because there had long been evidence of a cosmological constant), and those who knew how hard the problem was and had paid attention to Weinberg's prediction.  Speaking as one of the latter, my attitude toward the vacuum energy was not so much expectation as fear, and the hope that the evidence would go away, because a vacuum energy would bring the multiverse into physics.

It is often said that the multiverse is not predictive, but in a very real sense the exact opposite is true.  In the last half century there has been no more surprising, or more important, discovery about the fundamental nature of our universe than the vacuum energy.  This was the prediction most worth making, and only the multiverse made it.  Of course, this is a special circumstance coming from the extreme discrepancy between other theories and the observed value, and for many other quantities it is unpredictive.

However, we do not get to decide how predictive the laws of nature are, how much is random or environmental and how much is fixed.  It is something that we have to discover.  Of course, if the answer is that we live in a less predictable universe, it will be much harder to know that this is right.  But we can figure it out.

To conclude this section, I will make a quasi-Bayesian estimate of the likelihood that there is a multiverse.   To establish a prior, I note that a multiverse is easy to make: it requires quantum mechanics and general relativity, and it requires that  the building blocks of spacetime can exist in many metastable states.  We do not know if this last  assumption is true.  It is true for the building blocks of ordinary matter, and it seems to be a natural corollary to getting physics from geometry.  So I will start with a prior of 50\%.  I will first update this with the fact that the observed cosmological constant is not enormous.  Now, if I consider only known theories, this pushes the odds of a multiverse close to 100\%.  But I have to allow for the possibility that the correct theory is still undiscovered, so I will be conservative and reduce the no-multiverse probability by a factor of two, to 25\%.  The second update is that the vacuum energy is not exactly zero.  By the same (conservative) logic, I reduce the no-multiverse probability to 12\%.  The final update is the fact that our outstanding candidate for a theory of quantum gravity, string theory, most likely predicts a multiverse.\footnote{With a few exceptions, those who argue in the opposite direction are  indulging in wishful thinking.  By the way, I count the prediction of a multiverse as a sixth unexpected success of string theory: it accounts for the small but nonzero cosmological constant.}  But again I will be conservative and take only a factor of two.  So this is my estimate for the likelihood that the multiverse exists: 94\%.\footnote{For those who find this calculation amusing, I ask you: how many of you expected a nonzero cosmological constant in 1997?  If not, perhaps you should be a bit more humble.  Like Dick Cheney pontificating on Iraq, past performance is an indicator of future results.}

Occam may be respected in some circles where Bayes is not, so let me also express this in terms of his razor, as in~\cite{TV}.   Occam charged for assumptions.   As a physicist, it  seems that physics from geometry, and uniqueness of dynamics, are a minimal count to the razor.   On the other hand, it has always amazed me that general relativity does not charge for volume, and quantum mechanics does not charge for branches: these are no count against the razor.  The way that physics creates a rich universe from simple laws is that simple assumptions leave to rich dynamics, and the multiverse may be more of the same.

This is not to say that the multiverse is on the same footing as the Higgs, or the Big Bang.  Probability 94\% is two sigma; two sigma effects do go away (though I  factored in the look-elsewhere effect, else I would get a number much closer to 1). 
The standard for the Higgs discovery was five sigma, 99.9999\%.

\section{The next 116 years}

I started by talking about the great barrier posed by the Planck scale.  If the physical laws that we directly see  are environmental rather than fixed, then we face a second barrier that is just as great.  You can disagree with my 94\% estimate, but there is no rational argument that a multiverse does not exist, or even that it is unlikely.  

Supposing that both barriers do exist, one might think that progress will be impossible.  Indeed, if one takes an overly rigid definition of science, then science will be useless by definition.  But one should not be so pessimistic, or so dismissive of science.  We have perhaps had two pieces of bad luck, but also one of good luck: string theory exists, and we have found it.  

So how to go forward?  Experimentalists must explore all avenues, and theorists should follow their own instincts.  Other speakers will talk about the ongoing program to connect string vacua to observation.  But for me there is a clear challenge: to complete string theory, the same way that Einstein finished GR, using the kind of tools  that have brought us to our current understanding.  Obviously it will take longer, it has the quantum as well as the relativity.  It may require concepts that we do not yet suspect.  But we will succeed. This subject is currently in an exciting state, with new ideas from quantum information theory~\cite{Pastawski:2015qua,kitaev} and the black hole information problem,~\cite{Almheiri:2012rt} et seq.

So what is my prediction for 2131?  It is that we will have figured out the theory of quantum gravity, and that it will be built upon string theory.  What and how much will it be able to predict I cannot say; how much can be predicted is something for us to discover.  But unification has always led to unexpected insights and predictions, and so I will be optimistic, 
and predict that  when we have crossed the greatest gap and completed the greatest unification, we will know that we are correct, and the result will be wonderful.  Of course I am working hard to reduce the interval, I want to know the answer.

\section*{Acknowledgments}

I am grateful to Richard Dawid and his collaborators Radin Dardashti,
George Ellis,  Dieter Lust,  Joseph Silk, and Karim Thebault, and to the Munich Center for Mathematical Philosophy, LMU Munich, and the
Arnold Sommerfeld Center for Theoretical Physics, LMU Munich, for sponsoring this unique and important meeting.  I thank Raphael Bousso, Lars Brink, Andre Linde, Dave Morrison, and Bill Zajc for  communications.
The work of J.P. was supported by NSF Grant PHY11-25915 (academic year) and PHY13-16748 (summer).

\end{document}